\documentclass[11pt,a4paper]{article}
\pdfoutput=1
\usepackage{jheppub}
\usepackage{dsfont}
\usepackage{amssymb,amsmath}
\usepackage{epsfig}
\usepackage{epstopdf}
\usepackage{latexsym}
\usepackage{graphicx}
\usepackage{subfigure}
\usepackage{booktabs}
\usepackage{bbm}

\def\x{{\hat{x}} }
\def\O{{\cal O} }

\def\A{{\cal A} }

\def\N{{\cal N} }
\def\S{{\cal S} }
\def\L{{\cal L} }

\def\H{{\cal H} }
\def\cM{{\cal M} }

\def\R{{\mathbb R}}
\def\T{{\mathbb T}}
\def\Z{{\mathbb Z}}
\def\dcut{{%
    \setbox0\hbox{D}%
    \rlap{\hbox to \wd0{\hss ~/ \hss}}\box0}}

\subheader{\begin{flushright}
UTTG-14-13\\
TCC-010-13
\end{flushright}}

\title{Holographic entanglement in a noncommutative gauge theory}

\author{Willy Fischler$^{a,b}$,~Arnab Kundu$^a$,~Sandipan Kundu$^{a,b}$}
\affiliation{$^a$Theory Group, Department of Physics, University of Texas, Austin, TX 78712}
\affiliation{$^b$Texas Cosmology Center, University of Texas, Austin, TX 78712}
\emailAdd{fischler@physics.utexas.edu}
\emailAdd{arnab@physics.utexas.edu}
\emailAdd{sandyk@physics.utexas.edu}

\abstract{In this article we investigate aspects of entanglement entropy and mutual information in a large-$N$ strongly coupled noncommutative gauge theory, both at zero and at finite temperature. Using the gauge-gravity duality and the Ryu-Takayanagi (RT) prescription, we adopt a scheme for defining spatial regions on such noncommutative geometries and subsequently compute the corresponding entanglement entropy. We observe that for regions which do not lie entirely in the noncommutative plane, the RT-prescription yields sensible results. In order to make sense of the divergence structure of the corresponding entanglement entropy, it is essential to introduce an additional cut-off in the theory. For regions which lie entirely in the noncommutative plane, the corresponding minimal area surfaces can only be defined at this cut-off and they have distinctly peculiar properties.}

\begin{document}

\maketitle
\flushbottom

\section{Introduction}

Noncommutativity is an intriguing concept that has repeatedly appeared in mathematics and physics. It was realized, perhaps first in \cite{Snyder:1946qz}, that quantum field theories can be defined on a noncommutative geometry where the position coordinates do not commute. Based on a conventional Lagrangian description, the primary ingredient to define such a theory is an associative but not necessarily commutative algebra. Perhaps the most familiar example of such an algebra is the commutation relation between the space-coordinates
\begin{eqnarray}
\left[x^i , x^j \right] = i \theta ^{ij} \ ,
\end{eqnarray}
where $\theta^{ij}$ is the noncommutativity tensor.

Hallmark of a noncommutative theory is the inherent non-locality associated with the description. One key motivation behind studying such noncommutative theories is that they come equipped with a natural infra-red (IR) ultraviolet (UV) connection which is a profound property of any theory of quantum gravity.  In string theory noncommutative theories do appear rather naturally, of which the BFSS\cite{Banks:1996vh} and the IKKT\cite{Ishibashi:1996xs} matrix theories have been widely studied in the literature. On the other hand, non-locality posits conceptual challenges even for {\it simple} quantum field theories. Thus it will be a fruitful exercise to understand the {\it simple} non-local quantum field theories, which may eventually be relevant for a theory of quantum gravity.

In the current article we will attempt to investigate aspects of entanglement entropy in a noncommutative field theory. We will consider the noncommutative version of the ${\cal N}=4$ super Yang-Mills theory, where we just need to replace every commutative product by the star-product
\begin{eqnarray}
(f \star g)(x) \equiv \left.   e^{\frac{i}{2}\theta^{ij} \frac{\partial}{\partial {y^i}} \frac{\partial}{\partial {z^j}}} f(y) g(z) \right |_{y=z=x} \ ,
\end{eqnarray}
where $f(y)$ and $g(z)$ are two arbitrary functions.

Before proceeding further, let us briefly introduce the notion of entanglement entropy. In a given quantum field theory defined on a manifold, we imagine  an entangling surface that divides the entire manifold in two sub-manifolds at a given instant in time. Let us denote the corresponding sub-systems by $A$ and $A^c$ respectively. Consequently, the total Hilbert space factorizes: $\H_{\rm tot} = \H_A \otimes \H_{A^c}$. Now we can define a reduced density matrix for the sub-system $A$ by tracing over the degrees of freedom in $A^c$: $\rho_A = {\rm tr}_{A^c} [\rho]$. Finally we can use the von Neumann definition $S_A = - {\rm tr} \left[\rho_A \log \rho_A\right]$, which defines the entanglement entropy.

The astute reader will immediately notice potential subtleties in defining such a quantity in a noncommutative field theory. One obvious issue is how to define a {\it sharp}  entangling surface in a fuzzy manifold, which is also tied to the issue whether in a noncommutative theory $\H_{\rm tot} = \H_A \otimes \H_{A^c}$ factorization is sensible. At this point we can appeal to Bohr's correspondence principle and declare that we can define  {\it an} entangling surface provided we promote the classical algebraic equation defining such a surface as a statement on the corresponding operator and its eigenvalues.\footnote{We have elaborated more on this issue in the following sections.} The issue is now to carry out a computation.

We will not attempt to perform a weakly coupled field theory computation. Instead, we will make a detour {\it via} the AdS/CFT correspondence\cite{Maldacena:1997re} to analyze the issue in a large $N$ noncommutative gauge theory. By virtue of the gauge-gravity duality, we will need to perform a classical gravity computation in a geometric background obtained in \cite{Hashimoto:1999ut, Maldacena:1999mh}. In gauge-gravity duality, entanglement entropy is calculated using the Ryu-Takayanagi (RT) formula proposed in \cite{Ryu:2006bv}. According to this proposal, entanglement entropy of a region $A$ is given by 
\begin{eqnarray}
S_A = \frac{{\rm Area}(\gamma_A)}{4 G_N^{(d+1)}} \ ,
\end{eqnarray}
where $G_N^{(d+1)}$ is the $(d+1)$-dimensional Newton's constant, $\gamma_A$ denotes the minimal area surface whose boundary coincides with the boundary of region $A$: $\partial A = \partial \gamma_A$.

In the case of a noncommutative gauge theory defined on the boundary of the bulk geometry, this raises another subtlety: As we have argued before, $\partial A$ does not necessarily have a well-defined meaning in a fuzzy geometry. Perhaps the simplest generalization is to make sense of the boundary curve as an operator, denoted by $\widehat{\partial A}$, and define spatial regions as bounds on the eigenvalues of this operator.\footnote{We will discuss this in detail in section 2.3.} Thus there is no straightforward way to construct an interpolating bulk minimal area surface for such a spatial region. We will assume that, since the bulk geometry is still a classical one, the corresponding minimal area surface with the classical boundary $\partial A$,  if exists,  provides {\it a definition} for entanglement entropy. Henceforth we will investigate this quantity, both at zero and at finite temperature. Perhaps most intriguingly, the RT-formula --- as applied within our scheme of the prescription for a subregion residing entirely on the noncommutative plane --- does allow us to define a sensible entanglement. However, the corresponding minimal area surfaces have rather peculiar properties.

Let us now mull over why entanglement entropy may be a potentially interesting observable for such a theory. First, note that it is non-trivial to define a gauge-invariant operator in such a theory\cite{Gross:2000ba} and subsequently it is subtle to compare the corresponding results with an ordinary theory. Entanglement entropy, modulo the above-mentioned issues, can be an interesting probe for such theories. In the large temperature limit, entanglement entropy reduces to thermal entropy which is still a gauge-invariant concept for noncommutative theories. Therefore we expect that the entanglement entropy is also a gauge-invariant observable. Furthermore, this exercise may also shed light on the role of an inherent non-locality on quantum entanglement. It is expected that entanglement entropy obeys a universal area-law\cite{Srednicki:1993im} for a local field theory with nearest neighbour interaction\footnote{Note that even for a gauge theory, where the interactions are not nearest neighbour type, such an area law may hold. Although there is no general proof of this statement.}, since quantum entanglement occurs primarily across the common boundary. In a non-local theory this may not necessarily be true. We will find in several examples that there is a violation of the area law below a certain length-scale; however we will further argue that it may not be meaningful to probe the theory below this scale.\footnote{We will discuss this in detail in a later section. For earlier related studies, see \cite{Barbon:2008ut}.}

The above feature is more prominent in the so called mutual information, which is a derived quantity that has some advantages over entanglement entropy. Mutual information between two disjoint, separated sub-systems $A$ and $B$ is defined as
\begin{eqnarray}
I(A, B) = S_A + S_B - S_{A \cup B} \ , \nonumber
\end{eqnarray}
where $S_Y$ denotes the entanglement entropy of the region $Y$. It can be proved under general considerations that mutual information is always bounded by the area of the boundary of $A$ and/or $B$\cite{PhysRevLett.100.070502}. It is, however, not immediately clear  whether in a gauge theory the analogue of this theorem can tolerate some non-locality.\footnote{We would like to thank Matthew Hastings for correspondences on this issue.} We will find that mutual information is always an area-worth quantity, when it is finite. This area-worth and finite behaviour can be violated below a certain length scale, where the entanglement entropy also deviates from an area-law. Furthermore, it undergoes the familiar\cite{MolinaVilaplana:2011xt, Fischler:2012uv} entanglement/disentanglement transition as observed in generic large $N$ gauge theories. Perhaps more importantly, we will also show that mutual information is the right quantity to compare the results of the noncommutative theory  with the corresponding commutative one.

This article is divided in the following sections: We begin with a brief review of the noncommutative gauge theory and its holographic dual geometry. Within the same section we then discuss one possible way to define sub-regions of various possible shapes on a noncommutative geometry. We then discuss features of entanglement entropy and mutual information for a ``rectangular strip" geometry both at vanishing and at finite temperature in section 3 and 4 respectively. After this, we discuss some features of entanglement for more general shapes: a commutative cylinder in section 5 and a noncommutative cylinder in section 6. Finally we conclude in section 7.

\section{Noncommutative Yang-Mills}

\subsection{Holographic dual}

A mild form of non-locality can be realized by considering noncommutative gauge theories. In this section, we will consider a four-dimensional maximally supersymmetric SU(N) super Yang-Mills theory on a spacetime $\R_{\theta}^2\times \R^{1+1}$, where noncommutativity parameter is non-zero only in the $\R_{\theta}^2$ plane. $\R_{\theta}^2$ plane is defined by a Moyal algebra 
\begin{equation}
 [x^2, x^3]=i \theta \ . \label{comrel}
\end{equation}
At large N and  strong 't Hooft coupling, a holographic description of this theory can be given which, in the string frame, reads\cite{Hashimoto:1999ut, Maldacena:1999mh}\footnote{Note that the simplest way to understand why this corresponds to a noncommutative gauge theory is to consider an open string in the corresponding background, which yields the commutation relation in (\ref{comrel})\cite{Seiberg:1999vs}.}
\begin{align}
& ds^2= R^2\left[-u^2 f(u) dt^2 + u^2 dx_1^2+ u^2 h(u) \left(dx_2^2+dx_3^2\right)+ \frac{du^2}{u^2 f(u)}+d\Omega_5^2\right] \ , \label{MR} \\
& B_{23}=R^2 a^2 u^4 h(u) \ , \qquad e^{2 \Phi}=g_s ^2 h(u) \ ,\\
& F_{0123u}= \frac{4R^4}{g_s} u^3 h(u) \ , \qquad C_{01}=\frac{R^2 a^2}{g_s} u^4 \ . 
\end{align}
Here $R$ denotes the radius of curvature of the background geometry, $x^1$ and $t$ represent the $\R^{1,1}$-directions, whereas $\{x^2,x^3\}$ represents the $\R_\theta^2$-plane. The radial coordinate is denoted by $u$; the ultraviolet (boundary) is located at $u\to u_b$, where $u_b$ is a momentum cut-off that is taken to be large. Also, $g_s$ denotes the string coupling which is related to the radius of the geometry {\it via}, $R^4= 4\pi g_s N \alpha'^2$, where $\alpha'$ is the string tension. Finally, $d\Omega_5^2$ denotes the metric on an unit $5$-sphere.

Note that, the background in (\ref{MR}) is also characterized by two functions, denoted by $f(u)$ and $h(u)$ respectively. These functions are explicitly given by
\begin{equation}
f(u) = 1 - \left(\frac{u_H}{u}\right)^4 \ , \quad  h(u)=\frac{1}{1+ a^4 u^4} \ .
\end{equation}
The function $f(u)$ represents the existence of a black hole in the geometry and $h(u)$ bears the signature that the dual gauge theory is noncommutative. Here $u_H$ denotes the location of the event horizon and $a$ is related to the noncommutativity parameter through: $a=\lambda^{1/4} \sqrt{\theta}$, where $\lambda$ is the 't Hooft coupling defined as $\lambda = 4 \pi g_s$. The parameter $a$ can be thought of as the ``renormalized" noncommutativity at strong coupling, since this is the parameter that will enter in every holographic computation.

Before proceeding further, a few comments are in order: First, note that when there is no black hole present in the background, {\it i.e.}~$u_H=0$, the infrared limit of the geometry is obtained by sending $u \to 0$. In this limit, we recover an AdS-space. On the other hand if we send $u_b \to \infty$, $h(u) \to 0$ and thus the geometry degenerates. Hence we need to impose $u_b <\infty$.

Also note that the background in (\ref{MR}) can be simply obtained by a chain of T-duality transformations on the familiar AdS-Schwarzschild$\times S^5$-background. The non-trivial B-field and the dilaton are generated as a consequence of this chain of duality transformations. Hence we can view the $\{x^2, x^3\}$-directions as a $2$-torus $\T_\theta^2 \cong \R_\theta^2/ \Z_2$. The strict limit of $u_b \to \infty$ can also be viewed as the degeneration of this $2$-torus.

\subsection{Regime of validity}

We can trust the supergravity solution only when the scalar curvature of the background is small compare to $\sim 1/ \alpha' = 1/l_s^2 $, where $l_s$ is the string length. This leads to the condition
\begin{equation}
a u_b \gg \frac{2}{\sqrt{\lambda}} \ ,
\end{equation}
which is trivially satisfied for large 't Hooft coupling. The UV cut-off $u_b$ can also be thought as the momentum cut-off. Another constraint comes from the fact that proper distance --- as measured by the metric in (\ref{MR}) --- of a coordinate distance $l$ along $\R_\theta^2$ or $\T_\theta^2$ is larger than $l_s$:
\begin{equation}
\frac{l}{a}\gg \frac{(u_b a)^{1/2}}{\lambda^{1/4}} \ .
\end{equation}Â¥
Therefore, if $\epsilon$ is the short distance cutoff then
\begin{equation}
\frac{\epsilon}{a}\sim \frac{(u_b a)^{1/2}}{\lambda^{1/4}} \ .
\end{equation}Â¥
Finally, as explained before, we also need to impose the constraint that the $\R_\theta^2$ or $\T_\theta^2$ --- as measured by the metric in (\ref{MR}) --- does not degenerate. To sharpen this constraint, let us introduce the following dimensionless ``cut-off"
\begin{eqnarray}
\alpha = a u_b \ . \label{cut2}
\end{eqnarray}
Later we will observe that this ``cut-off" does play an important role in the divergence structure of entanglement entropy.

\subsection{Noncommutativity and entanglement entropy}\label{ncee}

Let us begin with an elementary discussion of defining regions using entangling surfaces in a noncommutative geometry. We will consider a $(3+1)$-dimensional noncommutative spacetime of the form $\R^{1,1} \otimes \R_\theta^2$ or $\R^{1,1} \otimes \T_\theta^2$. Let us take $x_2$ and $x_3$ to be the noncommutative directions. In analogy with quantum mechanics, we should treat $x_2$ and $x_3$ as operators with
\begin{equation}
\left[\x_2, \x_3\right]= i  \theta \ .
\end{equation}
In the commutative case, an entangling surface can be defined as an algebraic equation
\begin{eqnarray}
F\left(x_1 , x_2, x_3 \right) = 0 \ ,
\end{eqnarray}
which defines two sub-regions denoted by $A$ and $B$ respectively
\begin{align}
A& = \left\{ \left(x_1, x_2, x_3\right) \left. \right| F\left(x_1, x_2, x_3\right) \le 0 \right\} \ , \\
B&= \left\{ \left(x_1, x_2, x_3\right) \left. \right| F\left(x_1, x_2, x_3\right) \ge 0 \right\} \ .
\end{align}
In the noncommutative case, following the correspondence principle the function $F$ should be promoted to an operator 
\begin{equation}
F \rightarrow \hat{F}(\x_1, \x_2, \x_3) = \hat{F}(x_1, \x_2, \x_3) \ .
\end{equation}
The eigenstates of $\hat{F}(x_1, \x_2, \x_3)$ form a complete basis
\begin{equation}
\hat{F}(x_1, \x_2, \x_3)|F\rangle= F |F\rangle \ .
\end{equation}
Now the system can be divided in two sub-regions in a unique way:
\begin{align}
A& = \{|F\rangle\left. \right| F \le 0\} \\
B& = \{|F\rangle \left.\right| F \ge 0\} \ .
\end{align}
For a given function $F$, this division is unique.

Now, we can ask the following general question: if we choose a function $F(x_1, x_2, x_3)$ on the boundary and use the bulk geometry to calculate the RT-entropy
\begin{equation}
S_{\rm RT}(F) = \frac{{\rm Area}(\gamma_F)}{4 G_N^{(4+1)}} \qquad \partial \gamma_F = F \ ,
\end{equation}
what does it correspond to when the boundary theory is defined on a noncommutative background?

\subsubsection{$F(x_1, x_2, x_3)=F(x_1, x_2)$}

This is the simplest case. In this case, the answer is straight forward since the sub-regions have a boundary $\partial A$ or $\partial B$ that lies entirely in the commutative submanifold: $\partial A$, $\partial B$ $\in$ $\cM_{\rm com}$, where $\cM_{\rm com} \subset \R \otimes \R_\theta^2$. In the boundary theory, we should look at $\hat{F}(x_1, \x_2)$. Now, $\hat{F}$ and $\x_2$ commute and we can work in the $|x_2\rangle$ basis:
\begin{equation}
\hat{F}(x_1, \x_2)|x_2\rangle= F (x_1, x_2)|x_2\rangle \ . \label{commgen}
\end{equation}
Therefore, the RT-entropy $S_{\rm RT}(F)$ gives the geometric entropy between spatial regions $A$ and $B$:
\begin{align}
A& = \left \{ \left(x_1, x_2\right) \left. \right| F(x_1, x_2) \le 0 \right \} \\
B& = \left \{ \left(x_1, x_2\right) \left. \right| F(x_1, x_2) \ge 0 \right \} \ .
\end{align}

\subsubsection{General case}

When, $F$ is a function of both $x_2$ and $x_3$, then in the boundary theory the function $F(x_1, x_2, x_3)$ does not have a well-defined meaning since we cannot draw a sharp curve in the fuzzy $\R_\theta^2$-plane. In the bulk theory however, we still can calculate $S_{\rm RT}(F)$ following the usual bulk prescription. 

Although an entangling surface cannot be drawn in the general case, for any function $F(x_1, x_2, x_3)$ in the bulk theory, there is a unique decomposition in the boundary theory:
\begin{align}
A& =  \{|F\rangle \left. \right|  F \le 0\} \\
B& = \{|F\rangle \left. \right| F \ge 0\} \ .
\end{align}
Thus in this case, the operator $\hat{F}$ defines the quantum analogue of an entangling surface. It is therefore interesting to investigate whether $S_{\rm RT}(F)$ is the entanglement entropy between subsystems $A$ and $B$: $S_{\rm RT}(F) = S(A)$.

\section{Infinite rectangular strip}

In this section we will investigate one particular example for which we can easily perform the explicit computations. In particular, we will take the ``infinite strip" geometry. The background in (\ref{MR}) is written in the string frame, hence we will use the generalized RT-formula for the 10-dimensional geometry with a varying dilation
\begin{equation}
S_A=\frac{1}{4G_N^{(10)}}\int d^8 \sigma e^{-2\Phi}\sqrt{G_{\rm ind}^{(8)}}=\frac{\A}{4G_N^{(5)}} \ ,
\end{equation}
where $G_N^{(10)}=8\pi^6 \alpha'^4$, $\sigma$ parametrizes the worldvolume of the minimal surface and the $5$-dimensional Newton's constant $G_N^{(5)}$ is proportional to $G_N^{(10)}$ up to a volume factor.

\subsection{Commutative rectangular strip}
Let us choose
\begin{eqnarray}
X \equiv x^1 \in \left[ - \frac{l}{2}, \frac{l}{2}\right] \ , \quad x^2,x^3  \in \left[ - \frac{L}{2}, \frac{L}{2}\right] \ ,
\end{eqnarray}
with $L \rightarrow \infty$. The extremal surface is translationally invariant along $x^2,x^3$ and the area of the extremal surface (in the Einstein frame) is simply given by
\begin{equation} \label{areagen}
\A= \frac{L^2 R^3}{g_s^2}\int du u^{3}\sqrt{ X'^2+ \frac{1}{u^4 f(u)}} \ .
\end{equation}
The above expression coincides with the corresponding expression in a pure AdS-Schwarzschild background. Hence, the entanglement entropy is the same as that for the $\N=4$ SYM case.

\subsection{Non-commutative rectangular strip}

We will begin with the case when there is no black hole in the geometry, {\it i.e.}~by setting $f(u) =1$. Let us now compute the entanglement entropy for an infinite strip specified by
\begin{eqnarray}
X \equiv x^2 \in \left[ - \frac{l}{2}, \frac{l}{2}\right] \ , \quad x^1,x^3  \in \left[ - \frac{L}{2}, \frac{L}{2}\right] \ .
\end{eqnarray}
with $L \rightarrow \infty$. The corresponding extremal surface is translationally invariant along $x^1, x^3$ and the profile of the surface in the bulk is given by $X(u)$. Area of this surface (in the Einstein frame) is given by
\begin{equation} \label{areagen}
\A= \frac{L^2 R^3}{g_s^2}\int du u^{3}\sqrt{ X'^2+ \frac{1}{u^4 h(u)}} \ .
\end{equation}
One crucial comment is in order: The bulk metric in (\ref{MR}) is anisotropic in $\{x^2, x^3\}$ and $x^1$-directions. Naively, it looks like the physical distance along $x_2$ or $x_3$ should be $\sim l\sqrt{h(u_b)}$ at a given cut-off $u=u_b$. However, the relevant metric one should use to compute physical distances is not the bulk metric, but the open-string metric. One can check that the open-string metric for the background in (\ref{MR}) is still AdS and hence the physical distance is the same as the coordinate distance.

Now we will go ahead and obtain the equation of motion using action (\ref{areagen}) 
\begin{align} \label{eomgen}
\frac{dX}{du}= \pm \frac{u_c^{3}}{ u^{5} \sqrt{\left(1- \frac{u_c^{6}}{u^{6}}\right)h(u)}} \ ,
\end{align}
where, $u_c$ is an integral of motion and $u=u_c$ represents the point of closest approach of the extremal surface. Such surfaces have two branches\footnote{For convenience, we will refer to these as the U-shaped profiles. We also note our results here are consistent with \cite{Barbon:2008ut}.}, joined smoothly at $(u=u_c, X=0, X' \to \infty)$ and $u_c$ can be determined using the boundary conditions:
\begin{equation}
X(u_b) = \pm\frac{l}{2} \ ,
\end{equation}
which leads to
\begin{align} \label{lengen}
\frac{l}{2}=&\int_{u_c}^{u_b}\frac{u_c^{3}du}{ u^{5} \sqrt{\left(1- \frac{u_c^{6}}{u^{6}}\right)h(u)}}
\end{align}
and finally the area functional is
\begin{align} \label{ncyma}
\A = \frac{2L^2 R^3}{g_s^2}\int_{u_c}^{u_b}\frac{u du}{ \sqrt{\left(1- \frac{u_c^{6}}{u^{6}}\right)h(u)}} \ .
\end{align}
\begin{figure}[!]
\centering
\includegraphics[width=0.6\textwidth]{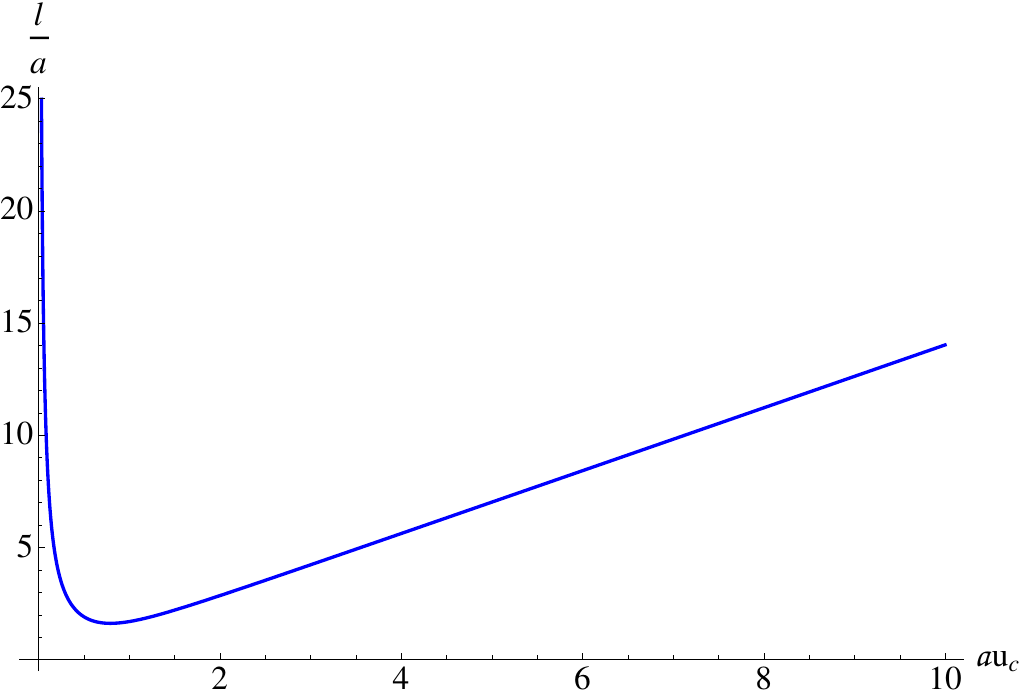}
\caption{Variation of $l$ with $u_c$ for U-shaped profiles. The curve has a minimum at $u_c a\sim0.8, \frac{l}{a}\sim1.6$. For $\frac{l}{a}>1.6$, two U-shaped solutions exist for each $l$. For $\frac{l}{a}<1.6$, U-shaped solution does not exist.}
\label{luc}
\end{figure}
This area is divergent and using the UV-cutoff $u_b$, we can write
\begin{equation}
 \A=\A_{\rm div} +\A_{\rm finite} \ .
\end{equation}
For the case in hand we get
\begin{align}
\A_{\rm div} =  & \frac{2L^2 R^3 a^2}{g_s^2}\left( \frac{u_b^4}{4}+\frac{1}{2 a^4}\ln (a u_b)\right) \ , \label{areadiv1}
\end{align}
whereas $\A_{\rm finite}$ can be calculated from the equation (\ref{ncyma}). But before we proceed, a few comments are in order. The U-shaped profile exists only for $l\ge l_0\sim 1.6 a$. In fact for $l>l_0$, two U-shaped solutions exist for each $l$; however the solution with $u_c a < 0.8$ has smaller area.

There is another solution to the extremal surface equation: $u=u_b$, which does not penetrate the bulk at all. For this class of solutions, we get the following divergence structure
\begin{equation}
\A_{\rm deg}= \frac{L^2 l R^3 u_b^3}{g_s^2} \ . \label{areadiv2}
\end{equation}
Comparing the leading behaviours in (\ref{areadiv1}) and (\ref{areadiv2}), we can conclude that the $u = u_b$ is the minimum area solution only when $l< l_c$, where,
\begin{equation}
 l_c=\frac{a^2 u_b}{2}+\frac{1}{ a^2 u_b^3}\ln (a u_b)\sim \frac{a^2 u_b}{2} \ .
\end{equation}
Let us now comment on why $l< l_c$ is not a sensible regime, although the entanglement entropy for this particular case is still a perfectly well-defined quantity even for $l<l_c$. It is well-known that in a noncommutative theory with noncommutativity parameter $\theta$ and a momentum cut-off $\Lambda$, the transverse (to the momentum) direction stretches to $\sim \theta \Lambda$. In our case, the parameter $(a^2)$ plays the role of a renormalized noncommutativity parameter and the bulk radial cut-off $u_b$ plays the role of a momentum cut-off. Thus for a momentum $p_3 \sim u_b$ along $x^3$, the uncertainty in the transverse direction becomes $\Delta x^2 \sim a^2 u_b$. Thus for the noncommutative theory at this energy scale, though formally entanglement entropy is still a well-defined quantity, it does not make sense to probe below the length scale $l_c \sim a^2 u_b$. Hence existence of $u=u_b$ solution, indicates that it is only sensible to consider length scales $l > l_c$. Here we should also note that we are interested in the regime $u_ba>>1$ where one can check that $l_c>>l_0$ and hence we always have U-shaped solutions.

From the bulk point of view, the $u=u_b$ solutions are also rather peculiar. First of all, note that the corresponding RT-surface does not probe the bulk geometry at all. Moreover, unlike the other familiar well-defined RT-surfaces, which obey the boundary condition that $X' \to 0$ as $u \to u_b$, this particular kind of surface obeys $X' \to {\cal O}(\infty)$ at the boundary. It is important to note that such solutions, for which $X'$ diverges at the boundary, result in an apparent volume dependence for entanglement entropy. This subsequently results in a divergent mutual information. In a later section we will observe a stronger presence of similar behaviour.

Thus we will discard the solutions $u=u_b$ henceforth.\footnote{In other words, we consider length scales that are greater than $l_c$.} Let us now focus on the divergence structure obtained in (\ref{areadiv1}). Naively, it seems that the entanglement entropy has a quartic divergence in $u_b$, as opposed to the quadratic divergence in the ordinary Yang-Mills case. This is evidently counter-intuitive. We cannot fit more than one degree of freedom inside one Moyal cell, which implies that, at the very least, noncommutativity should not worsen the divergence in a quantum field theory.

We will now argue that there is one way to reconcile with the above expectation. Let us recall that we introduced a dimensionless ``cut-off" $\alpha$ in (\ref{cut2}). Using this additional ``cut-off" we can rewrite the  divergent piece as
\begin{eqnarray}
\A_{\rm div}= \frac{L^2 R^3}{2 g_s^2} \alpha^2 u_b^2 + \frac{L^2 R^3}{g_s^2 a^2} \log \alpha \ ,
\end{eqnarray}
which can be re-interpreted as having a {\it familiar} quadratic divergence in $u_b$. Evidently this comes at the cost of having to introduce an additional scheme-dependent quantity $\alpha$. Note however, that taking a naive $a \to 0$ limit does not reproduce the known result for ordinary Yang-Mills theory. At this level, we can make a curious observation: Given the background in (\ref{MR}), we can first perform an RT-computation and then take the $a\to 0$ limit. Alternatively, we can also take the $a\to 0$ limit and then perform the RT-computation. These two processes do not yield the same result. This observation holds true for the finite part of the entanglement entropy as well, which we will discuss now.

\begin{figure}[!]
\centering
\includegraphics[width=0.7\textwidth]{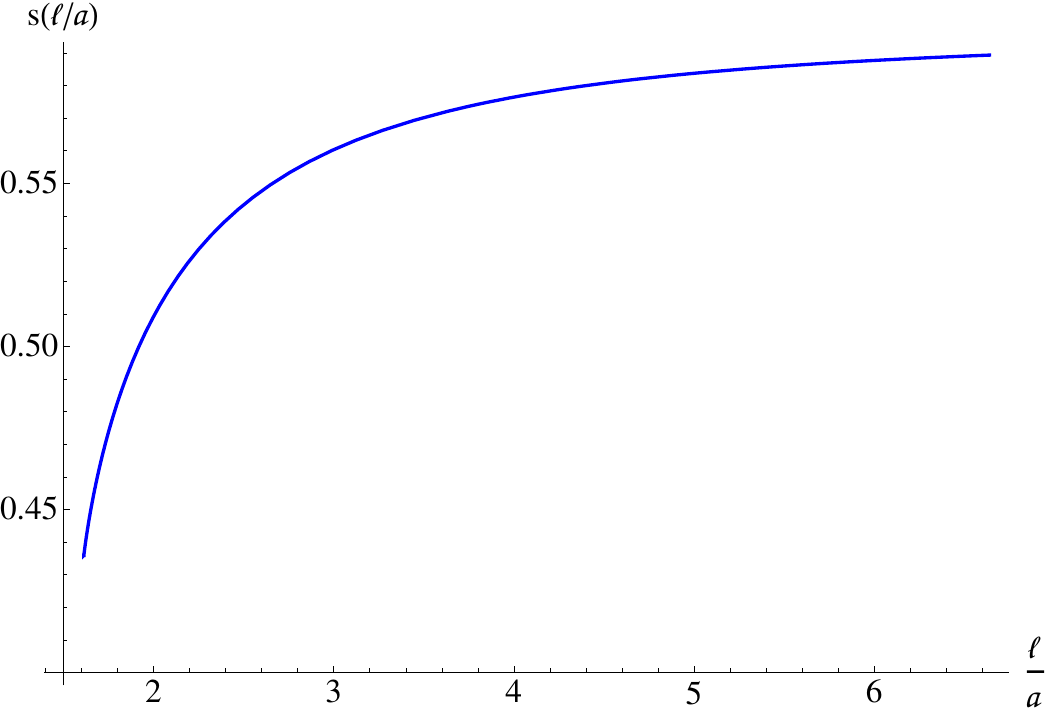}
\caption{Variation of $s(l/a)$ with $l/a$.}
\label{eeT0}
\end{figure}

We can schematically write the entanglement entropy ($l > l_c$) as
\begin{align}
S= S_{\rm div} + \frac{N^2}{2\pi}\left(\frac{L^2}{a^2}\right)s\left(\frac{l}{a}\right) \ .
\end{align}
Where $s(l/a)$ is a monotonically increasing finite-valued function of $l/a$ and for $l/a\rightarrow \infty$, it approaches $s(l/a)\rightarrow 0.5966$. Note that, in the ordinary large $N$ Yang-Mills case, the finite part of the entanglement entropy approaches zero as the length of the rectangular strip asymptotes to infinity. This is clearly not the case here. This again makes the comparison between the commutative and the ordinary Yang-Mills theories subtle. We will show that this subtlety is absent in the mutual information. Meanwhile, the functional behaviour of $s(l/a)$ can be evaluated numerically, which is shown in fig.~\ref{eeT0}.

Before proceeding further in discussing aspects of mutual information, let us ponder over a key aspect of entanglement entropy in this case. It is straightforward to observe that the divergent piece in (\ref{areadiv1}) is independent of the length of the interval $l$ and thus one can define a finite quantity derived from the entanglement entropy as follows
\begin{eqnarray}
C(l) = l \partial_l S (l/a) = \frac{N^2}{2\pi}\left(\frac{L^2}{a^2}\right) l \partial_l s(l/a) \ . \label{cfn}
\end{eqnarray}
In an $(1+1)$-dim CFT, this defines a central charge function that can be shown to be monotonically decreasing under an RG-flow\cite{Casini:2004bw} and hence measures the number of degrees of freedom: $l \partial_l C(l) \le 0$. It is straightforward to check that the above inequality is satisfied by (\ref{cfn}) since the curve in fig.~\ref{eeT0} is a concave one.

In general the above result follows from three criteria: (i) Lorentz invariance, (ii) unitarity and (iii) strong subadditivity of entanglement entropy. In our construction, the full Lorentz invariance is broken, $SO(3,1) \to SO(1,1) \times SO(2)$, where $\{t, x^1\}$ has the $SO(1,1)$ symmetry and $\{x_2,x_3\}$ has the $SO(2)$ symmetry. Thus it is non-trivial that the inequality $l \partial_l C(l) \le 0$ is still satisfied, since there is no ``effective" Lorentz symmetry in the $\{t, x^2\}$-plane to protect it.

\subsection{Mutual information}

\begin{figure}[!]
\centering
\includegraphics[width=0.7\textwidth]{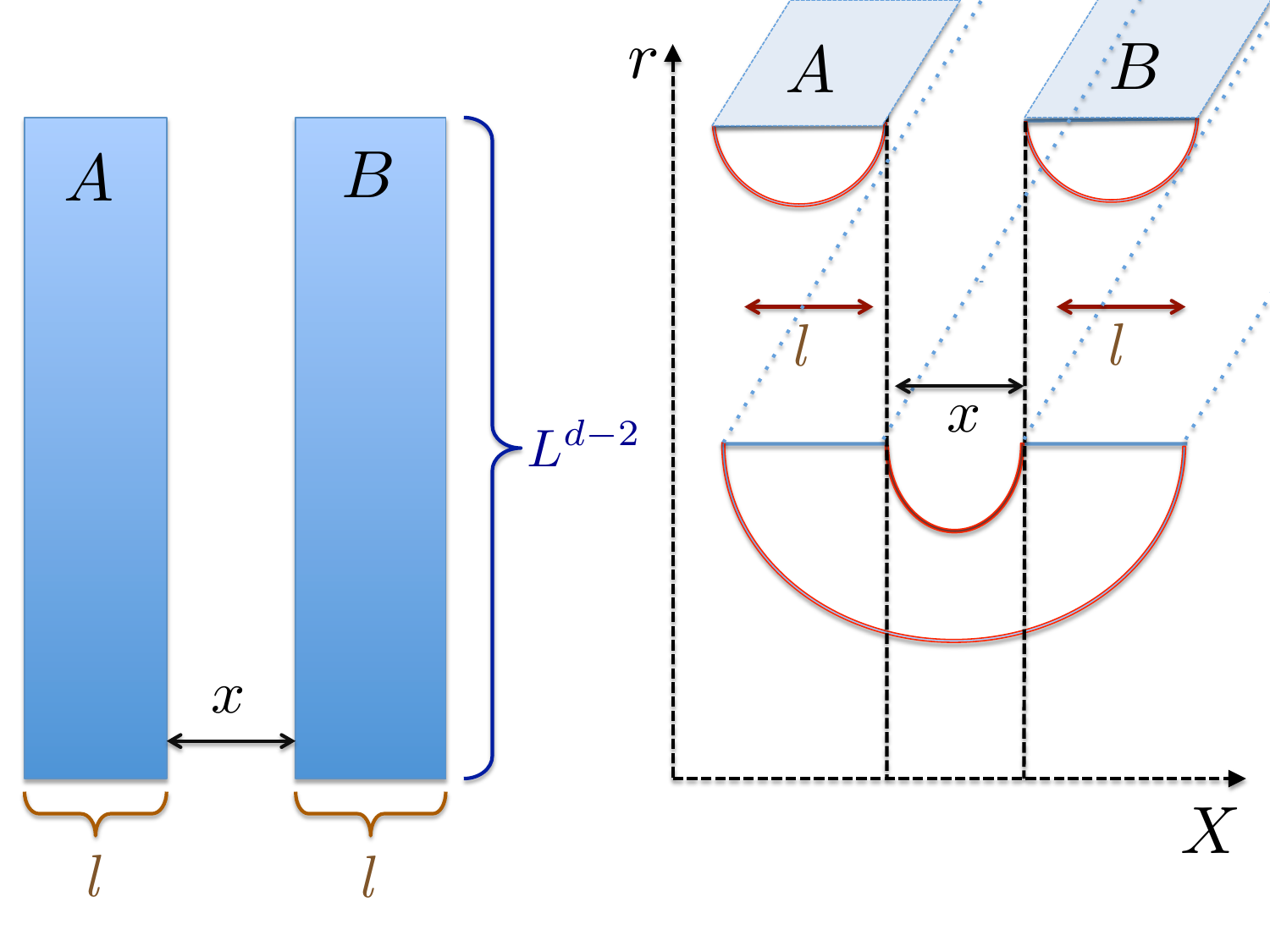}
\caption{A diagrammatic representation of the two rectangular strips which are used to analyze the mutual information. Here $X \equiv x^2$ and $r \equiv u$. Clearly, to compute the entanglement entropy of the region $A\cup B$, we have two choices for the minimal area surface.}
\label{recshape}
\end{figure}

Let us now move on to discuss mutual information. To define this quantity we need to consider two ``rectangular strips" each of width $l$ separated by a distance $x$ along the $x^2$-direction. For a visual rendition of the set-up, see fig.~\ref{recshape}. 
The corresponding mutual information between the sub-systems $A$ and $B$ is defined as
\begin{eqnarray}
I(A, B) = S_A + S_B - S_{A \cup B} \ , 
\end{eqnarray}
where $S_Y$ denotes the entanglement entropy of the region $Y$.

It is demonstrated in fig.~\ref{recshape} that for the computation of entanglement entropy of the region $A\cup B$, we have two candidates for the corresponding minimal area surface. This gives rise to an interesting ``phase transition" for mutual information, which has been discussed in details in \cite{Fischler:2012uv}. Here we will observe a similar physics. In this case mutual information is given by,\footnote{We consider $x,l>l_c$, as we have already mentioned. It can be checked easily that for $x$ and/or $l<l_c$, mutual information is divergent.}
\begin{eqnarray} \label{miT0}
I(A,B) & = & \frac{N^2}{2\pi}\left(\frac{L^2}{a^2}\right)\left[2s\left(\frac{l}{a}\right)-s\left(\frac{x}{a}\right)-s\left(\frac{2l+x}{a}\right)\right] , \quad x/l \le \beta \ , \\
           & = & 0 \ , \quad x/l >\beta \ ,
\end{eqnarray}
where $\beta$ depends on the non-commutative parameter $a$. The corresponding ``phase diagram" has been shown in fig.\ref{miT0}(a). Note that $\beta$ approaches the commutative result $0.732$ for large $x/a$. A couple of comments are in order: First, we note that mutual information again picks out the finite part of entanglement entropy. Second, it seems possible to recover the results for ordinary Yang-Mills theory by setting $a \to 0$ (see fig.~\ref{miT0}(b)), which is not the case for entanglement entropy.

\begin{figure}[!]
\centering
\includegraphics[width=0.7\textwidth]{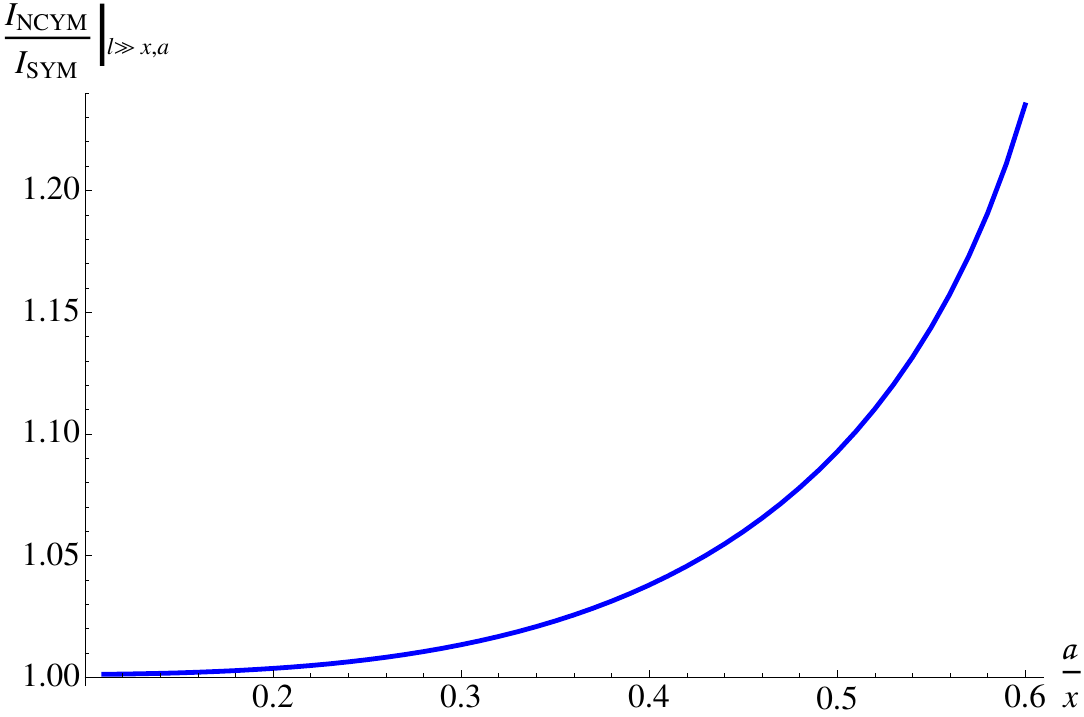}
\caption{Variation of mutual information (at large $l$ limit) with $a/x$, indicating an increase of correlations as we increase the noncommutativity of the theory. Also note that in the $a\to 0$ we recover the known result for pure SYM theory.}
\label{ratio_mi}
\end{figure}

To sharpen the latter statement, let us define the following quantity
\begin{equation}
 \left[\frac{I(A,B)_{\rm NCYM}}{I(A,B)_{\rm SYM}}\right]_{l\gg x,a} \ ,
\end{equation}
where ``NCYM" stands for non-commutative Yang-Mills and ``SYM" stands for super Yang-Mills. which we have plotted in figure \ref{ratio_mi}.
\begin{figure}[h!]
\begin{center}
\subfigure[] {\includegraphics[angle=0,
width=0.48\textwidth]{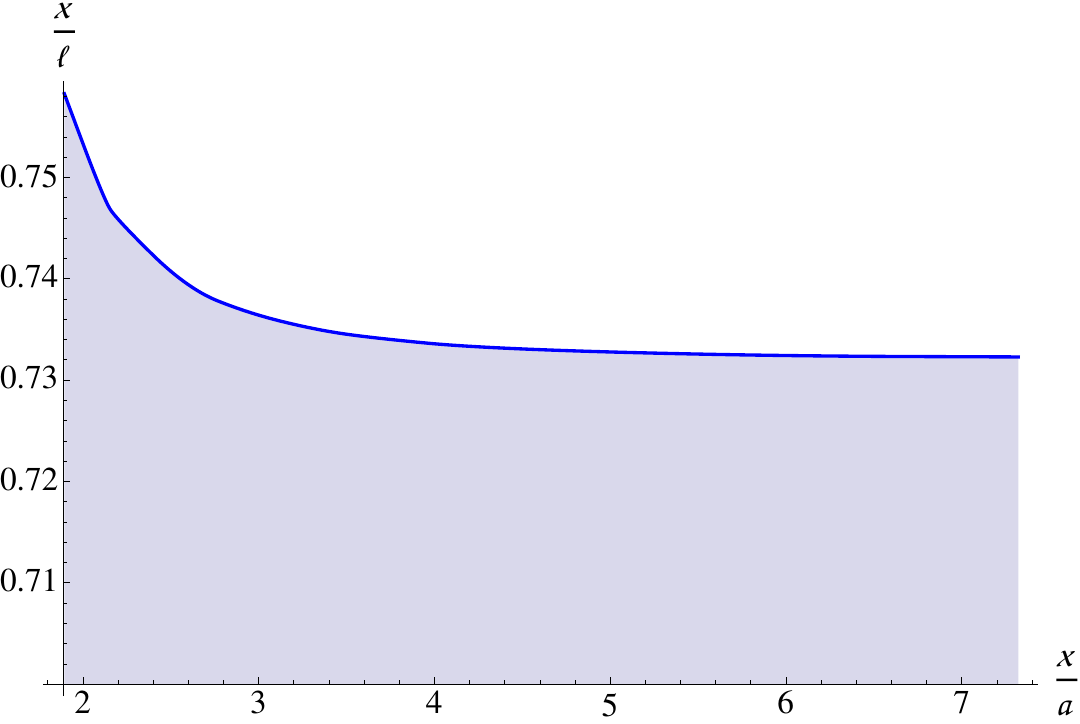} }
 \subfigure[] {\includegraphics[angle=0,
width=0.48\textwidth]{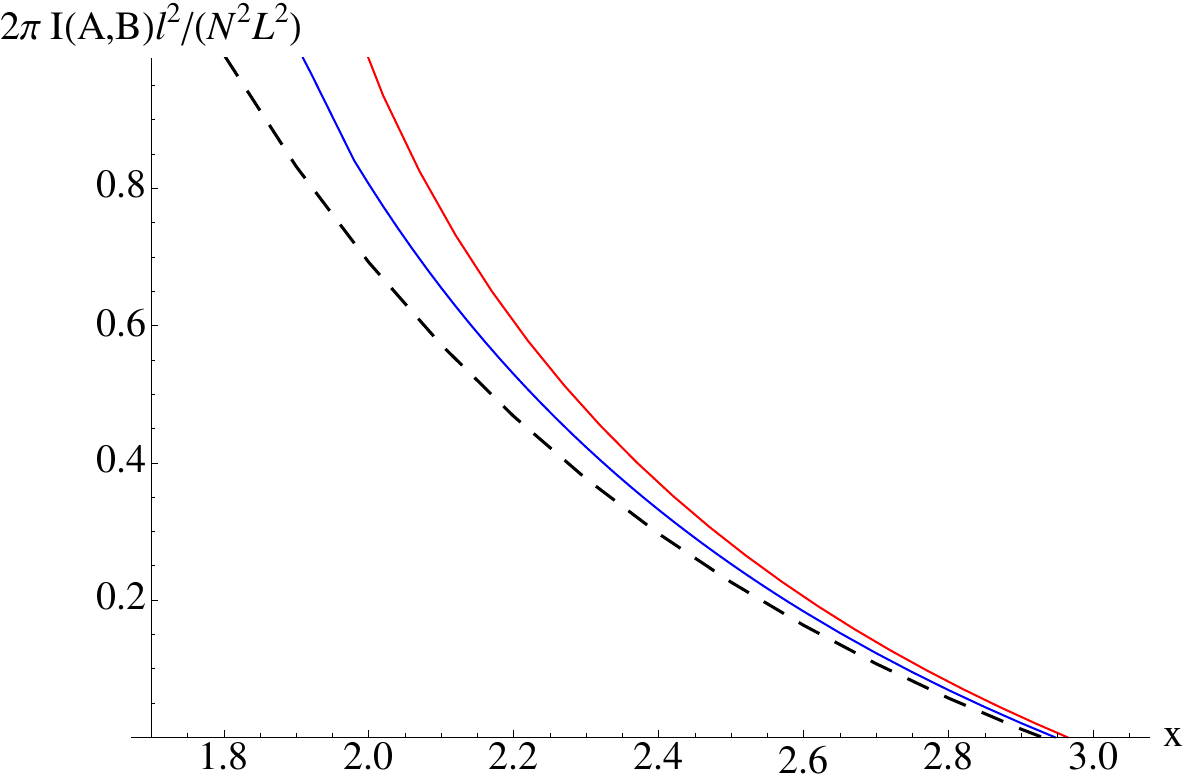} }\caption{Panel (a): 2-dimensional parameter space for the (3+1)-dimensional NCYM boundary theory. The mutual informational is non-zero only in the blue shaded region. Panel (b): Variation of $I(A,B)$ with $x$ for $l=4$. Blue solid line is for $a=1$ and red solid line is for $a=1.2$. Dashed black line shows the corresponding mutual information for the commutative case.}
\label{miT0}
\end{center}
\end{figure}
Since mutual information encodes {\it all possible} correlations, fig.~\ref{ratio_mi} implies that noncommutativity introduces more correlations between two sub-systems of the full system.

\section{Introducing Finite temperature}

We will now discuss the physics at finite temperature. To this end, we will now consider the geometry in (\ref{MR}) with $u_H \not = 0$. After Euclidean continuation and periodically identifying the time-direction, the corresponding temperature of the background can be obtained to be
\begin{eqnarray}
T = \frac{u_H}{\pi} \ .
\end{eqnarray}
Here we will confine ourselves to discuss the ``rectangular strip" geometry that we have been discussing so far.

\subsection{Entanglement entropy of an infinite rectangular strip}

Before we go ahead and calculate the entanglement entropy of a non-commutative infinite rectangular strip specified by
\begin{eqnarray}
X \equiv x^2 \in \left[ - \frac{l}{2}, \frac{l}{2}\right] \ , \quad x^1,x^3  \in \left[ - \frac{L}{2}, \frac{L}{2}\right] ,\qquad L \rightarrow \infty\ ,
\end{eqnarray}
let us discuss the nature of extremal surfaces for this particular geometric shape. 
\begin{figure}[!]
\centering
\includegraphics[trim = 10mm 40mm 10mm 0mm, clip,width=\textwidth]{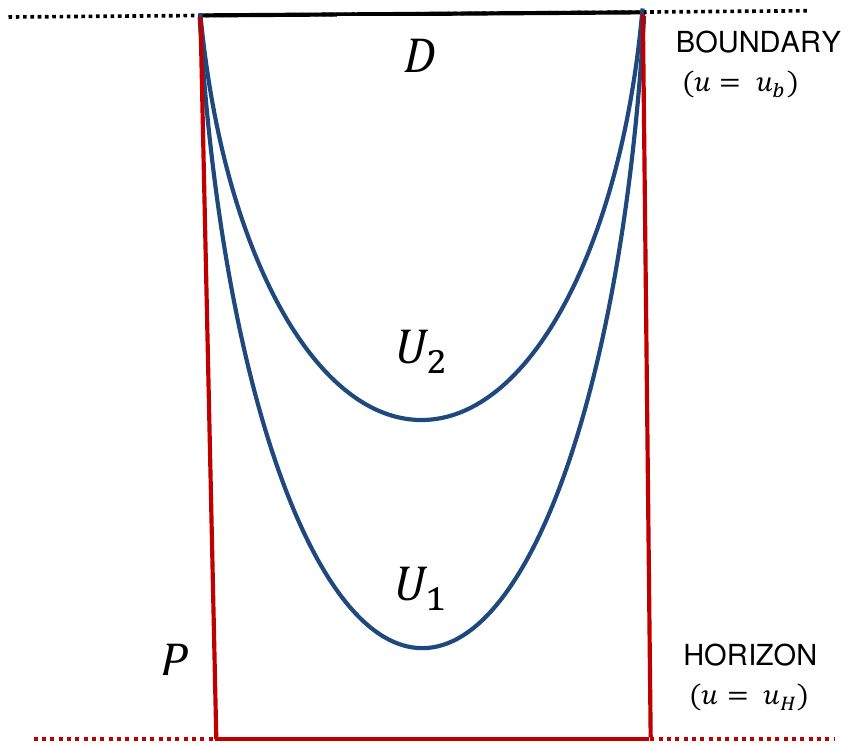}
\caption{All possible extremal surfaces for an infinite rectangular strip. There are four solutions: two U-shaped ($U_1, U_2$), one parallel ($P$) and one constant $(D)$ solutions.}
\label{extsur}
\end{figure}
Typically, for a given length $l$, there are four solutions (see figure \ref{extsur}). Just like the zero temperature case, there are two U-shaped solutions ($U_1, U_2$) when $l>l_0(aT)$.\footnote{Presence of the second U-shaped solution probably is the bulk reflection of the boundary UV/IR connection. It is interesting to note that in the limit $a\rightarrow 0$, solution $U_2$ approaches the constant solution $D$. When the noncommutativity is turned off completely,  $U_2$ coincides with $D$ and we are left with only one U-shaped solution.} The one that goes deeper into the bulk has smaller area
\begin{equation}
\A(U_1)\le \A(U_2) \ .
\end{equation}
For $l<l_0(aT)$, U-shaped solutions do not exist. It can be shown that at large temperature $(aT>>1)$
\begin{equation}\label{l0}
l_0(aT)\approx 2a^2 u_H=2\pi a^2 T \ .
\end{equation}
Similar to the commutative case, there exists a parallel solution ($P$) which has larger area
\begin{equation}
\A(U_1)\le \A(P)
\end{equation}
and hence it can only be important when $l<l_0(aT)$. As before, there is also a constant solution($D$): $u=u_b$, that does not go inside the bulk at all. This solution becomes important only when $l<l_c$
\begin{equation}
\A(D)< \A(U_1),\A(U_2),\A(P) \ ,
\end{equation}
where, we will show later that $l_c\sim \frac{1}{2} a^2 u_b$ does not depend on temperature. As we argued before, presence of this solution indicates that it does not make sense to probe below $l_c$.

From the above discussion, it is clear that we have two length scales: $l_0(aT)$ and $l_c$. At zero temperature, we saw that $l_0\sim 1.6 a$ is a fundamental length scale that comes from the noncommutative nature of the space-time. Whereas, $l_c$  is an effective cut-off that appears only after we introduce a momentum cut-off $u_b$. At zero temperature, $l_c$ is the relevant cut-off scale because $l_c>> l_0$. However, at high temperature, $l_0(aT)$ grows linearly with temperature (\ref{l0}) which probably indicates that the space-time  becomes fuzzier at finite temperature. 
\begin{figure}[!]
\centering
\includegraphics[width=0.7\textwidth]{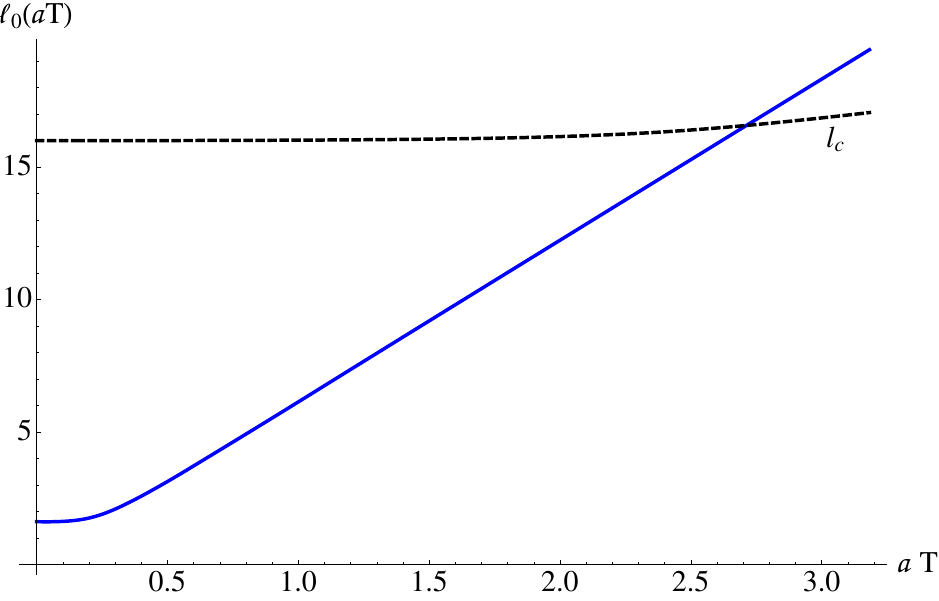}
\caption{Variation of $l_0(aT)$ (solid blue curve) with $aT$. Dashed black line corresponds to some $l_c$ for a particular momentum cut-off $u_b$. At low temperature $l_c>>l_0(aT)$. At sufficiently high temperature $l_0(aT)$ can be larger than $l_c$.}
\label{lc}
\end{figure}
It is possible to have $l_0(aT)>l_c$ only at sufficiently high temperature $(T\gtrsim\frac{u_b}{4\pi})$; this feature is schematically represented in fig.~\ref{lc}. In this regime, there seems  to be a second order phase transition of entanglement entropy at $l=l_0(aT)\sim 2\pi a^2 T$ from $U_1$ to $P$. However, this phase transition can be an artifact of having temperature $T$ close to the momentum cut off $u_b$; this should be investigated more carefully in future. At temperature $T$, average momentum along any direction $p^i\sim T$ and uncertainty along noncommutative directions become $\Delta x^{2,3}\sim a^2 T$. Hence one perhaps can argue that this phase transition is physically irrelevant because at very high temperature ($T\sim u_b$) it is not sensible to probe below $l\sim l_0(aT)$.  

\begin{figure}[!]
\centering
\includegraphics[width=0.7\textwidth]{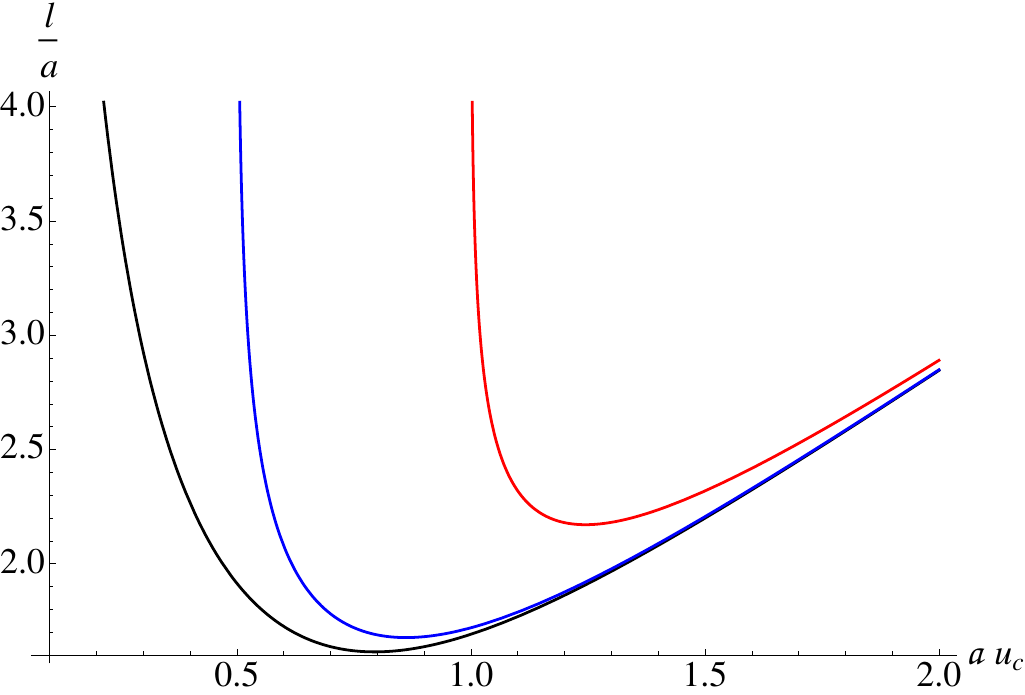}
\caption{Variation of $l$ with $u_c$ for u-shaped profiles at finite temperature. The black curve is for zero temperature, blue and red lines are for $u_h a=0.5$ and $u_h a=1$ respectively.}
\label{lucT}
\end{figure}

Now we will compute the area of the physically relevant U-shaped solutions. Proceeding as before, at finite temperature we obtain
\begin{align} \label{lT}
\frac{l}{2}=&\int_{u_c}^{\infty}\frac{u_c^{3}\sqrt{1+a^4 u^4}du}{ u^{5} \sqrt{\left(1- \frac{u_c^{6}}{u^{6}}\right)\left(1- \frac{u_H^{4}}{u^{4}}\right)}} \ ,
\end{align} 
and
\begin{align} \label{AT}
\A=\frac{2L^2 R^3}{g_s^2}\int_{u_c}^{\infty}\frac{u \sqrt{1+a^4 u^4}du}{ \sqrt{\left(1- \frac{u_c^{6}}{u^{6}}\right)\left(1- \frac{u_H^{4}}{u^{4}}\right)}} \ .
\end{align}
As we mentioned earlier, for any $l$ above some $l_0(a u_H)$, there are two U-shaped solutions; the one with smaller value of $a u_c$ corresponds to smaller area. For increasing values of $a u_H $, $l_0$ monotonically increases and so does $a u_c$. This feature is pictorially represented in fig.~\ref{lucT}.

Now, at finite temperature the divergence structure is slightly different:
\begin{align}
S_{\rm div}= \frac{N^2}{2\pi}\left(\frac{L^2}{a^2}\right)\left( \frac{u_b^2 \alpha^2 a^2}{2}+\left(1+\pi^4 a^4 T^4\right)\ln (\alpha)\right) \ ,
\end{align}
which seems to receive an additional cut-off dependent term at finite temperature. Note that this is rather unique, since usually finite temperature does not introduce additional cut-off dependence in ordinary quantum field theories.

\begin{figure}[!]
\centering
\includegraphics[width=0.8\textwidth]{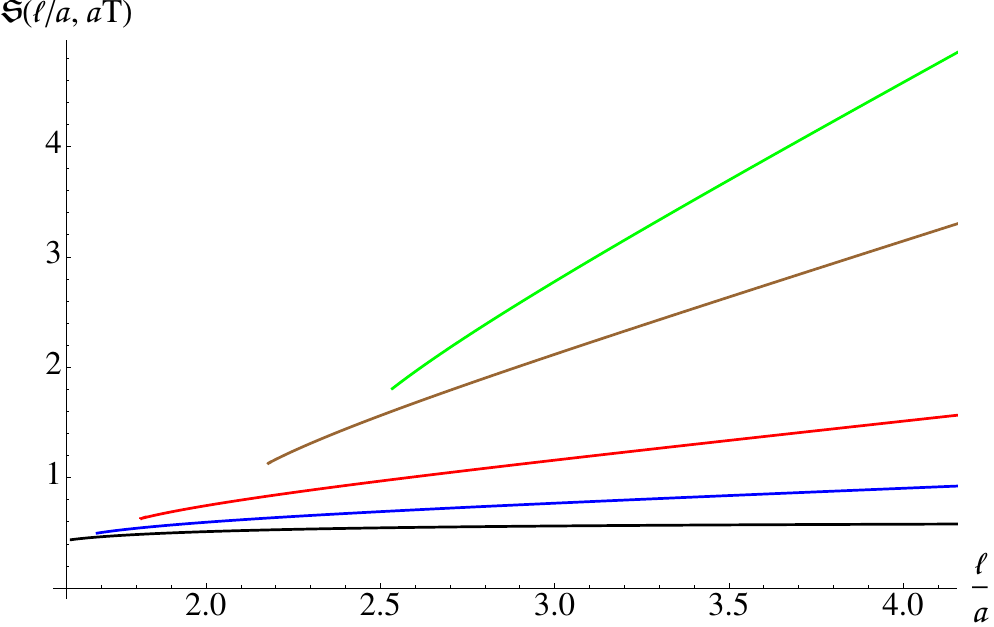}
\caption{Variation of $\S(l/a,Ta)$ with $l/a$ for $Ta=0$ (black), $Ta=\frac{0.5}{\pi}$ (blue), $Ta=\frac{0.7}{\pi}$ (red), $Ta=\frac{1}{\pi}$ (brown), $Ta=\frac{1.2}{\pi}$ (green). }
\label{ee_ncym_T}
\end{figure}

Similar to the zero temperature case, the constant solution(D) becomes important at smaller value of $l$ with a volume-worth of ``area"
\begin{equation}
\A_{\rm deg}= \frac{L^2 l R^3 u_b^3}{g_s^2}
\end{equation}
for
\begin{equation}
l< l_c=\frac{a^2 u_b}{2}+\left(\frac{1+\pi^4 a^4 T^4}{ a^2 u_b^3}\right)\ln (a u_b)\sim \frac{a^2 u_b}{2} \ .
\end{equation}
It is noteworthy that this $l_c$ does not receive strong contribution coming from the temperature. As argued before, we will discard such solutions. For $l>l_c$ (and $l_0$), we get
\begin{align}
S= \frac{N^2}{2\pi}\left(\frac{L^2}{a^2}\right)\left( \frac{u_b^2 \alpha^2}{2}+\left(1+\pi^4 a^4 T^4\right)\ln (\alpha)\right)+\frac{N^2}{2\pi}\left(\frac{L^2}{a^2}\right)\S\left(\frac{l}{a},T a\right) \ ,
\end{align}
where $\S$ is the finite part, which is pictorially represented in fig.~\ref{ee_ncym_T}. 
For large $lT$ the finite part of the entanglement entropy becomes linear in $l$, exactly like the commutative case. That means at large $lT$, the leading finite part of the entanglement entropy follows a volume law. At high temperatures, the most dominant contribution to the finite part of the entanglement entropy is expected to come from the near horizon part of the extremal surface \cite{Fischler:2012ca}, and it is given by
\begin{equation}
\S \sim \frac{\pi^2 N^2}{2}V T^3 \ . 
\end{equation}
which is independent of the noncommutativity parameter $a$.

\subsection{Mutual information}

\begin{figure}[!]
\centering
\includegraphics[width=0.7\textwidth]{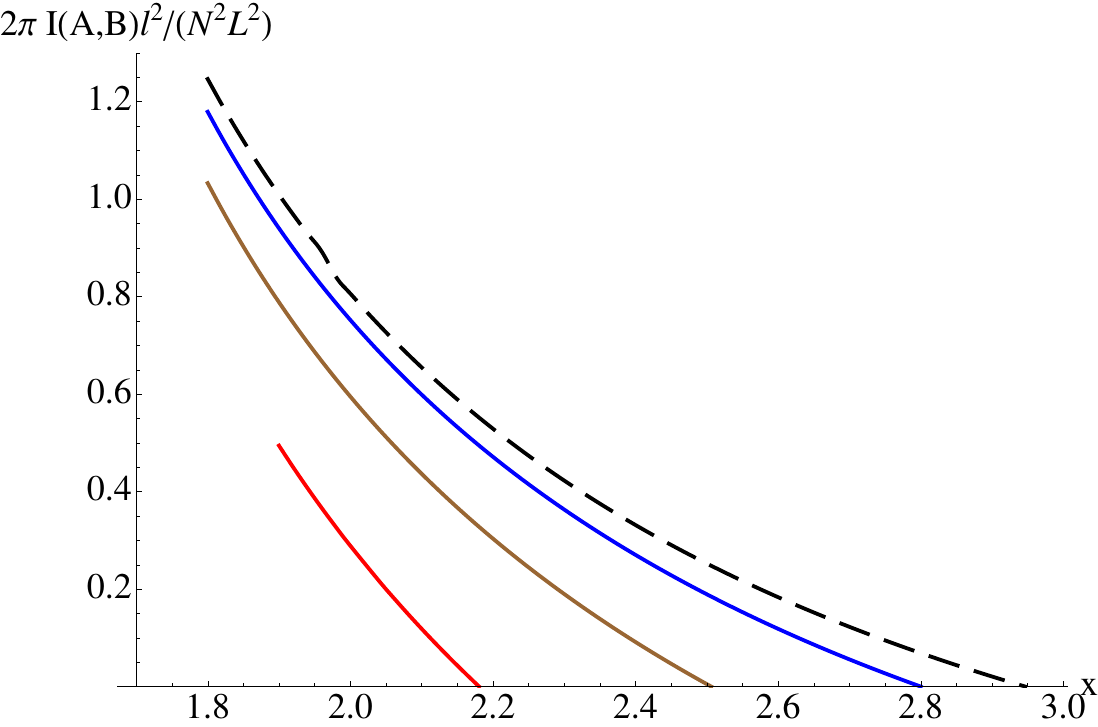}
\caption{Variation of $I(A,B)$ with $x$ for $l=4$ and $a=1$ for different temperatures. Blue solid line is for $T=\frac{0.1}{\pi}$,  brown line is for $T=\frac{0.15}{\pi}$ and red solid line is for $T=\frac{0.2}{\pi}$. Dashed black line shows the mutual information at $T=0$.}
\label{ncym_mi_T}
\end{figure}

\begin{figure}[h]
\centering
\includegraphics[width=0.8\textwidth]{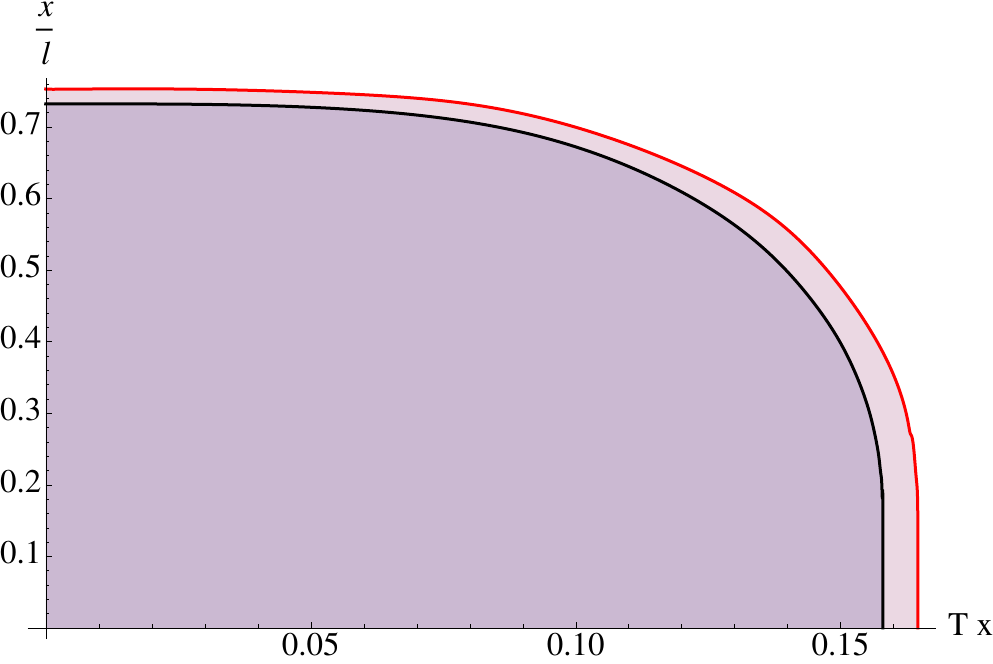}
\caption{2-dimensional parameter space for the (3+1)-dimensional boundary theory. Mutual informational is non-zero only in the shaded region. Mutual information is non-zero for the region below the black curve for the commutative case and below the red curve for the noncommutative case ($\frac{a}{x}=\frac{1}{2}$).}
\label{trans_ncym}
\end{figure}

Once again mutual information can be easily obtained from the entanglement entropy and for the configuration shown in figure \ref{recshape} it is given by
\begin{eqnarray} \label{miT}
I(A,B) & = & \frac{N^2}{2\pi}\left(\frac{L^2}{a^2}\right)\left[2\S\left(\frac{l}{a},T a\right)-\S\left(\frac{x}{a},T a\right)-\S\left(\frac{2l+x}{a}, T a\right)\right] , \quad x/l \le \beta_T \ \nonumber, \\
           & = & 0 \ , \quad x/l >\beta_T \ ,
\end{eqnarray}
where $\beta_T$ depends on the noncommutative parameter $a$ and temperature $T$.\footnote{We have assumed that $l,x>l_c$ and $T<<u_b$. One can again check that for $x$ and/or $l<l_c$, mutual information is divergent.} In fig.~\ref{ncym_mi_T} we have pictorially shown how mutual information behaves. As in the the case of vanishing temperature, above the length-scale $l_c$ mutual information is again a well-defined finite quantity which yields the ordinary Yang-Mills result in the limit $a \to 0$. Mutual information undergoes the expected\cite{Fischler:2012uv} disentangling transition and the corresponding ``phase diagram" is shown in fig.~\ref{trans_ncym}. As expected, we observe that a non-zero value of $a$ results in a larger region in the phase space where mutual information is non-zero. 
%

\section{More general shapes: commutative cylinder}

So far we have studied the simplest geometry, namely the ``infinite rectangular strip". The primary reason for this is technical simplicity. However, to gain intuition one needs to consider more general shapes as we will see below. However, many of the explicit computations become involved in such cases and we will not attempt a thorough analysis. Rather, we will focus on some qualitative features henceforth. For these purposes, we will consider the background at vanishing temperature.

\subsection{Entanglement entropy}

\begin{figure}[!]
\centering
\includegraphics[width=0.7\textwidth]{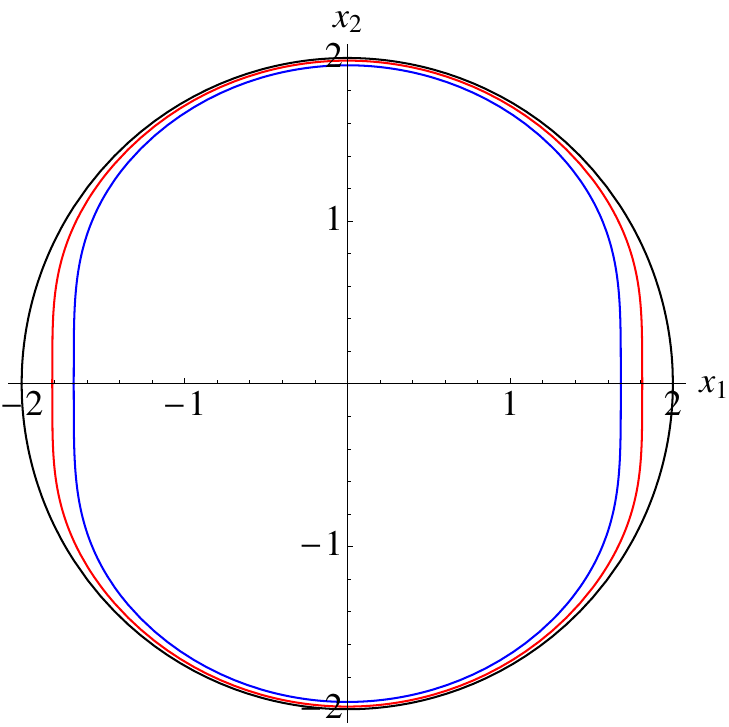}
\caption{Behavior of the extremal surface for the commutative cylinder near the boundary. Radius of the cylinder is $r a=2$ at the boundary (black curve). The red curve is at $ua=5$ and blue curve is for $ua=2$.}
\label{cylinder}
\end{figure}

Let us begin by considering a cylinder: a circle with radius $r$ in $x_1-x_2$ plane and length $L$ along $x_3$ and in the limit $L \to \infty$:
\begin{equation}
x_1=\pm \sqrt{r^2-x_2^2} \ .
\end{equation}
Note that the above curve falls under the general category discussed in (\ref{commgen}). The minimal area surface can be parametrized by $x_1 (u, x_2)$\footnote{For our purposes, it is particularly convenient to consider Cartesian coordinates. In the bulk geometry given in (\ref{MR}) representing a circle in the $\{x^1, x^2\}$-plane using a polar coordinate is inconvenient, since there is a non-trivial warp factor $h(u)$.}. Near the boundary, the extremal surface can be written in the following form:
\begin{equation}
F(u, x) \equiv x_1(u, x_2) = \pm \sqrt{r^2-x^2}+ F_1(u,x) \ ,
\end{equation}
with $F_1(u_b, x)\rightarrow 0$. The area functional is given by:
\begin{equation}
\A=\frac{R^3 L}{g_s^2} \int u dudx \sqrt{1+\left(a^4 u^4+1\right) \left(\frac{\partial F(u,x)}{\partial x}\right)^2+u^4 \left(\frac{\partial F(u,x)}{\partial u}\right)^2} \ .
\end{equation}
Corresponding equation of motion is given by
\begin{equation}
\left.\left.u \left(a^4 u^4+1\right) \frac{\partial }{\partial x}\left(\frac{1}{L_0} \frac{\partial F(u,x)}{\partial x}\right.\right)+\frac{\partial }{\partial u}\left(\frac{u^5}{L_0}\frac{\partial F(u,x)}{\partial u}\right.\right)=0 \ , \label{eomhard}
\end{equation}
where
\begin{equation}
L_0=\sqrt{1+\left(a^4 u^4+1\right) \left(\frac{\partial F(u,x)}{\partial x}\right)^2+u^4 \left(\frac{\partial F(u,x)}{\partial u}\right)^2} \ .
\end{equation}
The solution near the boundary is given by:
\begin{equation}
x_1(u, x_2) = F(u, x) = \pm \sqrt{r^2-x^2}\mp \frac{r^2 \log (u a)}{2 u^2 x^2 \sqrt{r^2-x^2}}+... 
\end{equation}
The sub-leading piece of the solution is a consequence of the fact that the bulk metric is anisotropic. Near the boundary the extremal surface can be represented as (see figure \ref{cylinder})
\begin{equation}
x_2^2-\sqrt{(r^2-x_1^2)^2-\frac{2 r^2 \log (u a)}{u^2 }}=0 \ .
\end{equation}
The curve above clearly demonstrates that the circle gets squashed along the $x^1$-direction as we move along the bulk radial direction. 

To obtain the corresponding entanglement entropy, we need to solve the equation in (\ref{eomhard}) numerically. Here we will not attempt so, instead let us focus on the divergence structure of the entanglement entropy which can be deduced from the leading order solution near the boundary. The divergent part of the entanglement entropy is given by\footnote{At finite temperature, similar to the infinite strip case, there will be an additional cut-off dependent term $\sim N^2 a^2 T^4 L r \ln \alpha$.}
\begin{equation}
S(A)_{\rm div} = \frac{N^2 L}{2\pi}\left[\alpha^2 r u_b^2+ \frac{\alpha^2}{r}\left(c_1 + c_2 \ln (\alpha)\right) +c_3 \frac{r}{a^2} \ln (\alpha)\right] \ .
\end{equation}
As we have argued before, the corresponding minimal area surface should have less area than a degenerate surface at $u=u_b$\footnote{Note that $u=u_b$ is not an exact solution of (\ref{eomhard}). We are taking this surface to estimate a possible lower bound on $r$.} with ``area" $\frac{N^2 L}{2\pi}u_b^3 \pi r^2$. This constraint sets a lower bound for the value for the radius of the cylinder
\begin{equation}
r\gtrsim \frac{1}{\pi}a^2 u_b=\frac{2}{\pi}l_c = r_c \ .
\end{equation}¥
This is reminiscent of the similar constraint we encountered earlier for the rectangular strip.

\subsection{Mutual information}

Here we will merely argue that mutual information is still a well-defined, finite and cut-off independent quantity above a minimal length-scale $\O(r_c)$. We can consider the case of three concentric circles of various radii and consider the two sub-regions as:
\begin{align}
&\text{Region} \, \, A = \{ (x_1, x_2) \left. \right| x_1^2+x_2^2 \le r_1^2 \} \ , \\
&\text{Region} \, \, B = \{ (x_1, x_2) \left. \right|  r_2^2 \le x_1^2+x_2^2 \le r_3^2 \} \ ,
\end{align}
where $r_1< r_2 < r_3$. In this case, the computation of $S(A\cup B)$ will be more involved since there are many candidate minimal area surfaces. However, it is easy to check that just like the $\N=4$ SYM case:
\begin{equation}
S(A \cup B)|_{\rm div}=S(A )|_{\rm div}+S( B)|_{\rm div} \ , 
\end{equation}
and hence
\begin{equation}
I(A,B)=\text{finite} \ .
\end{equation}¥
Thus it is again possible to construct a well-defined quantity derived from the entanglement entropy as evaluated using the RT-formula.

\section{More general shapes: noncommutative cylinder}

Now let us discuss potentially a more intriguing case. Let us consider constructing a circle in the $\{x^2, x^3\}$-plane and define the region $A$ by
\begin{eqnarray}
A = \{ (x_2, x_3) \left. \right| x_2^2 + x_3^2 \le r^2 \} \ . 
\end{eqnarray}
We also imagine that $x^1 \in \left[-L/2 , L/2 \right]$ with $L\to \infty$. Note that this case falls under the category where we {\it pretend} to draw a sharp curve in the otherwise fuzzy plane to investigate what the classical bulk RT-surface yields.

To proceed we define the corresponding polar coordinate in the plane {\it via}: $dx_2^2 + dx_3^2 = d\rho^2 + \rho^2 d\theta^2$. The bulk interpolating minimal area surface can now be parametrized by $\rho(u)$. With this, the area functional is given by
\begin{equation}
\A=\frac{2\pi R^3 L}{g_s^2} \int u^3 \rho(u) du \sqrt{\rho'(u)^2+\frac{1}{u^4 h(u)}} \ ,
\end{equation}
which yields the following equation of motion
\begin{equation}\label{eomnc}
\frac{d}{du}\left(\frac{u^3 \rho(u) \rho'(u)}{\L_0}\right)=u^3 \L_0 \ ,
\end{equation}
where
\begin{equation}
\L_0=\sqrt{\rho'(u)^2+\frac{1}{u^4 h(u)}} \ .
\end{equation}
Now it can be checked that 
\begin{equation}\label{good}
\rho(u)=r+ g(u) \qquad \text{with} \qquad  g(u a\rightarrow \infty)\rightarrow 0
\end{equation}
is not a solution, indicating there is no well-behaved solution for this case. Note that, had such a solution existed, it would mean $\rho'(u)/a^2 <<1$ at the boundary which is the hallmark of a well-behaved solution.

To investigate this case further we will find out the {\it best possible} solution to the equation of motion (\ref{eomnc}) that does not satisfy the condition (\ref{good}). This means that we allow $\rho'/a^2 \sim {\cal O}(1)$ at the boundary.\footnote{Note that $\rho'/a^2$ is a dimensionless quantity.}
\begin{figure}[!]
\centering
\includegraphics[width=0.95\textwidth]{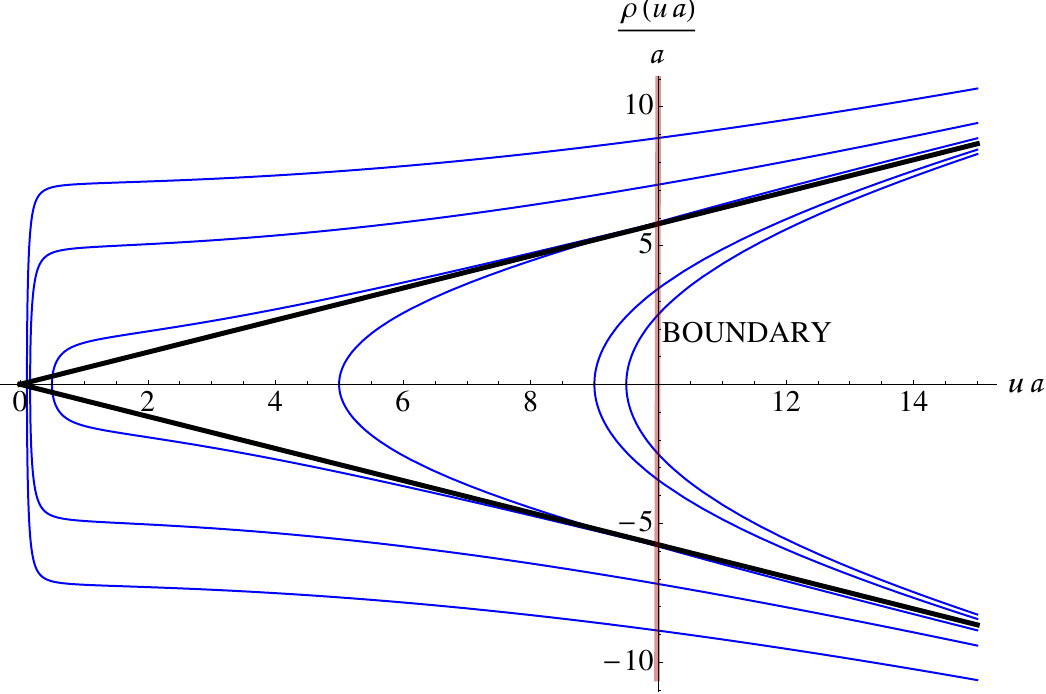}
\caption{Extremal surfaces for the noncommutative cylinder for different $r$ for $u_b a=10$. The black line corresponds to the leading behavior of the critical solution (\ref{critical}) and all the other U-shaped solutions asymptotically approach this solution as $(u_b  a)$ is taken to infinity. The vertical red line denotes the location of the boundary.}
\label{nccylms}
\end{figure}
With this condition, it can be shown that near the boundary ($ua>>1$) one solution of the equation of motion (\ref{eomnc}) behaves in the following way:
\begin{equation}\label{critical}
\rho_c(u)= \frac{a^2}{\sqrt{3}}u \left[1+\frac{1}{2a^4 u^4}-\frac{1}{8a^8 u^8}+ {\cal O} \left(\frac{1}{a^{12} u^{12}}\right) \right]\ ,
\end{equation}
Now imposing the boundary condition $\rho(u_b)=r$, for this solution we obtain
\begin{equation}
r_c=\frac{a^2}{\sqrt{3}}u_b \ .
\end{equation}
This solution yields the following divergence 
\begin{equation}
S(A)|_{\rm div}=N^2 L\left[\frac{2 a^4 u_b^5}{15}+\frac{u_b}{3}\right] =N^2 L\left[\frac{2 r_c^2 u_b^3}{5}+\frac{u_b}{3}\right]  \ .
\end{equation}
It looks like the entanglement entropy in this case has a volume divergence and is reminiscent of the volume divergence that we have encountered before while considering the rectangular strip or the commutative cylinder geometries earlier.

There are other U-shaped solutions for this case; we have showed them in figure \ref{nccylms}. These numerical solutions are obtained by solving the equation (\ref{eomnc}) with boundary conditions:
\begin{equation}
\rho(u_b)=r \qquad \text{and} \qquad \rho(u_c)=0 \ , \qquad \rho'(u_c)\to \infty \ ,
\end{equation}
where, $u=u_c$ is the closest approach point that depends on the radius $r$. One can check that for all these solutions at the boundary $\rho' (u)\sim {\cal O}(a^2)$ .

Note that we have previously encountered solutions with the following boundary behaviours: (i) $X' \to 0$ or (ii) $X' \to \infty$ as $u \to u_b$, where $X(u)$ represents the profile of the minimal area surface. For the class of solutions in (i), we obtain a familiar area-law behaviour for entanglement entropy and a finite mutual information. On the other hand, the degenerate minimal area surfaces in (ii), {\it e.g.}~the ones given by $u=u_b$ that do not probe the bulk geometry at all, result in divergent mutual information. The physics is qualitatively different for the noncommutative cylinder and the minimal area surfaces with the boundary behaviour $\rho' (u)\sim {\cal O}(a^2)$, probe the bulk geometry.

It is important to note that the solution in (\ref{critical}) is an attractor to all these solutions and hence in the limit $u_b a\rightarrow \infty$ all solutions correspond to the same radius in the boundary. Therefore, to make sense of the calculation of entanglement entropy and to allow ourselves to have various values of the radius, it is essential to introduce a cut-off $u_b a=\alpha$, which is large ($\alpha>>1$) but finite. 
\begin{figure}[!]
\centering
\includegraphics[width=0.8\textwidth]{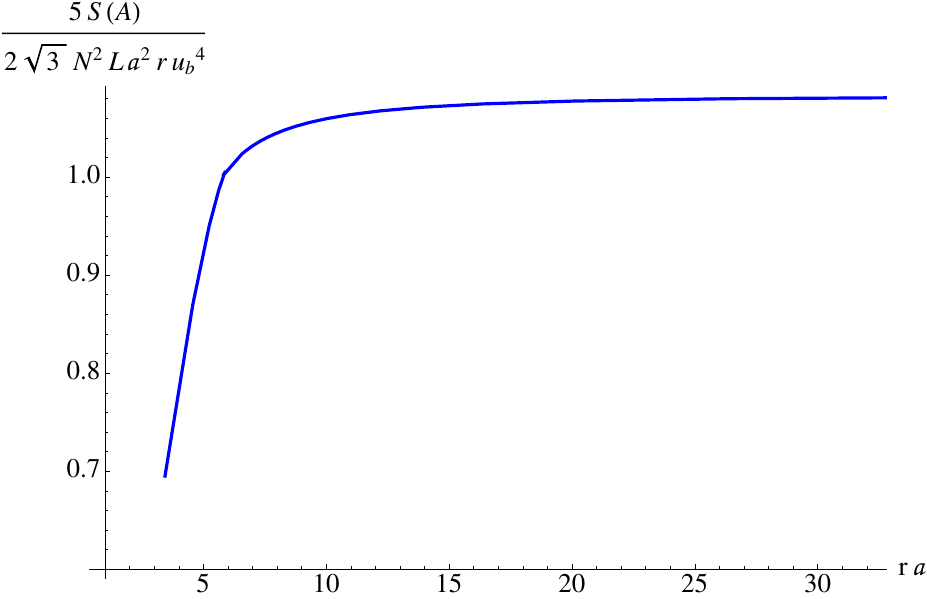}
\caption{Variation of entanglement entropy $S(A)$ for the noncommutative cylinder with $ra$ for $u_b a=10$. For $r>r_c$, $\frac{5S(A)}{2\sqrt{3}N^2 L u_b^4 a^2 r}\sim 1$, indicating an area law behavior.}
\label{nccylee}
\end{figure}

It is interesting to check the behavior of leading divergence for the U-shaped solutions. An analytic answer is no longer available; hence we have used numerical techniques to investigate the issue (see plot \ref{nccylee}). From figure \ref{nccylee}, it is clear that there is a transition at $r=r_c$. For $r>r_c$, we recover the familiar area law 
\begin{align}
S(A)|_{\rm div}=\frac{2\sqrt{3}}{5}N^2 L\left[ q(ra)\alpha^2 r u_b^2+\text{sub-leading terms}\right], \qquad r>r_c \ ,
\end{align}
where $q(ra)$ is an $\O(1)$ function of $r$ and it can be shown that $q(r_c a)=1$ and $q(ra\rightarrow \infty)\sim 1.08$. For $r<r_c$, figure \ref{nccylee} indicates a volume divergence,
\begin{align}
S(A)|_{\rm div}\sim N^2 L\left[p(ra)\frac{2 r^2 u_b^3}{5}+\text{sub-leading terms}\right]  , \qquad r<r_c \ ,
\end{align}
where $p(ra)$ is another $\O(1)$ function of $r$. The divergence structure heavily depends on $\rho' (u)$ at the boundary. Solution $\rho_c(u)$ in (\ref{critical}) is a critical solution with $\rho' (u)=a^2/\sqrt{3}$; for $r<r_c$, $\rho' (u_b)>a^2/\sqrt{3}$ (see figure \ref{nccylms}) and we have a volume divergence. This is also reminiscent of the volume behavior that we have
encountered before. For $r>r_c$ when $\rho' (u_b)<a^2/\sqrt{3}$ and we recover an area law for entanglement entropy. Interestingly, it can also be checked that for large $r$
\begin{equation}
\frac{S(A)|_{\rm cyl-noncom}}{S(A)|_{\rm cyl-com}}=\frac{4\pi \sqrt{3} q(ra)}{5}>1\ ,
\end{equation}
which implies that there is more entanglement for the noncommutative cylinder.

Before we conclude, a few comments are in order: let us try to connect these results with the discussion of section \ref{ncee} where
\begin{equation}
\hat{F}(x_1, \hat{x}_2, \hat{x}_3)= \hat{x}_2^2+\hat{x}_3^2-r^2
\end{equation}
with eigenvalues $F_n=(n+\frac{1}{2})a^2-r^2$.\footnote{Note that in our case, the parameter $a^2$ plays the role of renormalized noncommutivity parameter.} Thus, entanglement entropy $S(r)$ obtained by  tracing out all $|F_n\rangle$ with $F_n>0$ should be a step function of $r$ with step size $\delta r\sim a^2/2r$. However, RT-prescription provides us with an entanglement entropy $S(r)$ which is a continuous function of $r$ because the relevant length scale for these calculations is $r\sim u_b a^2$ and hence $( \delta r / r)  \ll 1$.

It is extremely difficult to compute mutual information on this noncommutative plane; however, strong dependence of the divergent part of the entanglement entropy on $\rho'(u)$  indicates that mutual information defined in the usual way, may not yield a cut-off independent behaviour even above $r=r_c$.

\section{Conclusions}

In this article we investigated aspects of quantum entanglement in a large $N$ noncommutative gauge theory using the AdS/CFT correspondence. We observed that the RT-formula allows us to obtain well-defined entanglement observables for a class of regions, which do not completely lie on the noncommutative plane. This comes at the cost of introducing an additional cut-off, which in the bulk geometry is realized as the degeneration of the noncommutative torus. The corresponding leading divergence structure in entanglement entropy is not altered by noncommutativity.

We have also observed that the role of this additional cut-off is more crucial if we want to define entanglement entropy, {\it via} the RT-prescription, for a region residing entirely on the noncommutative plane. In this case, the corresponding minimal area surfaces have distinctly peculiar properties, which may lead to a volume-law behaviour for entanglement entropy and also result in a divergent mutual information. It is interesting to note that if the violation of an area-law stems from the inherent non-locality of the theory, it is not clear why this violation is necessarily of volume-worth, rather than anything else bigger than the area.

Let us also note that in the large $N$ limit the noncommutative Yang-Mills theory does not differ from the ordinary Yang-Mills one for a number of observables, {\it e.g.}~the thermodynamics is identical in both cases\cite{Cai:1999aw}. This stems from the general result obtained in \cite{Bigatti:1999iz}, which states that all planar Feynman diagrams in a noncommutative Yang-Mills and an ordinary Yang-Mills theory are the same, unless there is an external momentum. In this article we have observed that entanglement entropy, as obtained using RT-formula actually receives a non-trivial contribution from the noncommutativity even at large $N$ and thus falls outside the class of observables for which the result in \cite{Bigatti:1999iz} holds. In the large temperature limit, however, we do recover the thermal entropy as expected.

There are various directions for future explorations. In this article we have only focussed on the divergent part of entanglement entropy for general shapes of the sub-regions. It will be interesting to analyze and understand the finite part of the entanglement entropy for such regions. For the ``rectangular strip", it will also be interesting to explore the possible second order phase transition of entanglement entropy at finite temperature, which we alluded to in section 4.

Let us conclude by saying that our analysis here does not involve any perturbative field theoretic computation. It will be interesting to consider analyzing entanglement entropy in a weakly coupled noncommutative field theory, or in a more elementary quantum mechanical set-up. It is well-known that noncommutative theories do play an important role in understanding physical phenomenon, such as the Quantum Hall Effect\cite{Susskind:2001fb}. Thus it may also be of direct physical relevance to entertain such questions, even though we may not learn anything directly related to issues in quantum gravity. We leave these issues for future explorations.

\section{Acknowledgements}

We would like to thank Elena Caceres, Jacques Distler, Matthew Hastings, Matthew Headrick, Juan Maldacena, Shiraz Minwalla and Mark van Raamsdonk for useful conversations and correspondences. AK would also like to thank the hospitality of the ICTS, Bangalore during the final stages of this work. This material is based on work supported by the National Science Foundation under Grant Number PHY-0906020 and by Texas Cosmology Center, which is supported by the College of Natural Sciences and the Department of Astronomy at the University of Texas at Austin and the McDonald Observatory. AK is also supported by a Simons postdoctoral fellowship awarded by the Simons Foundation.

\end{document}